\documentclass[twocolumn,preprintnumbers,amsmath,amssymb]{revtex4}
\usepackage{latexsym,amssymb,amsthm,amsmath,epsfig}
\usepackage{color}

\usepackage[hidelinks]{hyperref}

\begin{document}

\title{Unitarity-limited behavior of three-body collisions in a $p$-wave interacting Fermi gas}
\author{Muhammad Waseem$^{1,2,3}$}
\author{Jun Yoshida$^{1,2}$}
\author{Taketo Saito$^{1,2}$}
\author{Takashi Mukaiyama$^{3}$}
\affiliation{%
$^{1}$\mbox{Department of Engineering Science, University of Electro-Communications, Tokyo 182-8585, Japan}\\
$^{2}$\mbox{Institute for Laser Science,University of Electro-Communications, Chofugaoka, Chofu, Tokyo 182-8585, Japan}\\
$^{3}$\mbox{Graduate School of Engineering Science, Osaka University, Machikaneyama, Toyonaka, Osaka 560-8531 Japan}\\
}
\date{\today }

\begin{abstract}
We experimentally investigate the unitarity-limited behavior of the three-body loss near a $p$-wave Feshbach resonance in a single-component Fermi gas of $^6$Li atoms. At the unitarity limit, the three-body loss coefficient $L_{3}$ exhibits universality in the sense that it is independent of the interaction strength and follows the predicted temperature scaling law of $L_3 \propto T^{-2}$. When decreasing the interaction strength from the unitarity regime, the three-body loss coefficient as a function of the interaction strength and temperature can be described by the theory based on the association of an excited resonant quasibound state and its relaxation into a deep stable dimer by collision with a third atom in the framework of the standard Breit-Wigner theoretical approach. The results reported here are important to understand the properties of a resonant $p$-wave Fermi gas in the prospect of quantum few- and many-body physics.
\end{abstract}

\maketitle

In a system of particles with strong interactions, universal features often emerge. A prime example of such a system is $s$-wave unitary Fermi or Bose gases, where a tunable interaction parameter known as the scattering length diverges. Such a unitary gas exhibits universal properties that are independent of scattering length and depend only on the atomic density and temperature~\cite{hori, Nascimbene, Ku, solo,navon}. Furthermore, ultracold atomic gases at the unitarity limit attract interest from the context of few-body physics because unitary Bose gases have been confirmed to show the universal feature that the three-body loss coefficient depends only on the temperature~\cite{rem,zoran}.
The significant advancement in Bose gases with $s$-wave interactions has inspired studies toward fermionic gases with $p$-wave interactions~\cite{ohashi, vg, victor, Cheng, fed, sukjin, zhang, shunk, regal}. The universal features of few-body collisions in the atoms with $p$-wave interactions have been theoretically studied~\cite{esry1, esry2, suno, suno2, jona, jesper, jpd, lev, nishida,eric, maxim,zen2, nikolaj}.

Interactions among ultracold atoms across the $p$-wave Feshbach resonance are characterized by the scattering volume $V$, the collision energy $E$, and the effective range $k_{\rm e}$~\cite{ticknor, chevy, naka}. 
The scattering volume as a function of the external magnetic field is parameterized as $V=V_{\rm bg}[1-\Delta B/(B-B_{0})]$, where $V_{\rm bg}$, $\Delta B$, $B$, and $B_{0}$ are the background scattering volume, resonance width, external magnetic field, and  resonance position, respectively. The effective range can be safely assumed to be constant due to its very weak dependence on the magnetic field~\cite{lev,naka}. 
There are two distinct regimes of resonant $p$-wave Fermi gases, with weak and strong interactions~\cite{jesper,thy}.
A recent measurement of three-body loss coefficient near the $p$-wave Feshbach resonance in the lowest internal state of $^6$Li atoms confirmed the predicted scaling law of the three-body collision coefficient $L_{3}\propto V^{8/3}$~\cite{yoshida}.
This scaling solely holds in the weakly interacting regime, where scattering volume is small, and the temperature is low.
The deviation in the behavior of the three-body loss coefficient  from the weakly interacting regime has been observed in $^6$Li atoms~\cite{zhang}.
Since the first quantitative study of the three-body loss coefficient at the strongly-interacting regime~\cite{zhang}, further systematic investigation of the three-body loss coefficient has been awaited to fully understand the unitarity-limited behavior of three-body loss at the $p$-wave strongly-interacting regime, which has appealing interest in the context of few-body physics~\cite{suno,suno2,jpd}.

In this Rapid Communication, we report on the observation of the unitarity-limited behavior of the three-body loss coefficient in the vicinity of a $p$-wave Feshbach resonance in a spin-polarized Fermi gas of $^6$Li atoms. We experimentally show that at sufficiently large negative scattering volume or small magnetic field detuning from the resonance, the three-body loss coefficient reaches the limit imposed by the unitarity. 
We confirmed that at the unitarity limit, the three-body loss coefficient $L_3$ is not dependent on the interaction parameters and shows the temperature dependence of $L_3 \propto T^{-2}$ as predicted theoretically~\cite{suno,suno2}. We construct a simple theoretical model by taking into account the formation of a resonance quasibound state and its vibrational quenching by inelastic collision with other atoms based on the Breit-Wigner theoretical approach. Our theoretical approach successfully explains the magnetic-field and the temperature dependence of a three-body loss coefficient in a nonunitarity regime with strong interactions.

\begin{figure}[tbp]
\begin{tabular}{@{}cccc@{}}
\includegraphics[width=3.250 in]{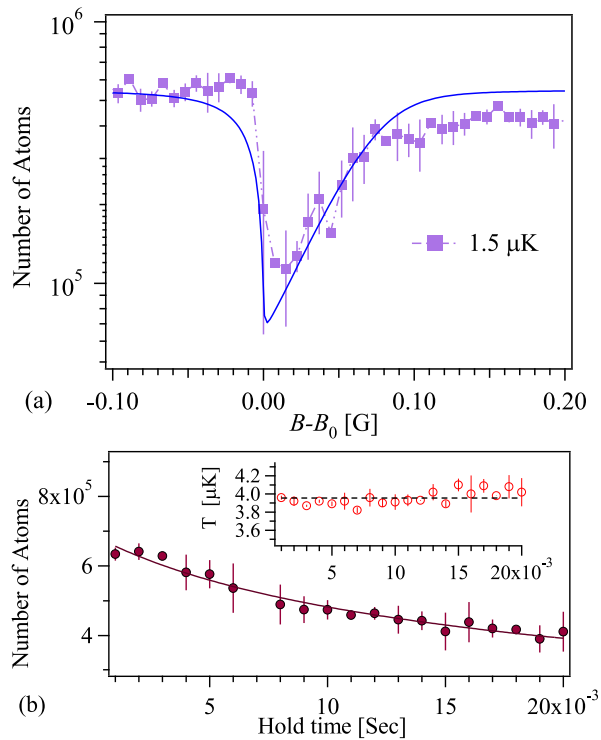} \\
\end{tabular}
\caption{(a) A plot of a remaining number of atoms after holding 20 ms at various magnetic field detunings $B-B_{0}$ at an atomic temperature of 1.5~$\mu$K. The sharp edge corresponds to the resonance position $B_0$ indicated by zero point. The solid curve is obtained using Eq.~(\ref{k3}) in the solution of rate equation~(\ref{rate}) as explained in text. (b) Typical time evolution of the number of atoms near resonance. Solid curve is a fitting to the solution of Eq.~(\ref{rate}) to extract three-body loss constant $L_{3}$. The inset shows the temperature evolution over hold times. Vertical error bars represent statistical errors due to atom number fluctuations.}
\label{fig1}
\end{figure}

Our experiment begins by trapping $^{6}$Li atoms in the lowest hyperfine ground state $\left\vert 1 \right\rangle \equiv \left\vert F=1/2,m_F=1/2 \right\rangle$, using a single beam optical dipole trap, as described in detail elsewhere~\cite{naka, yoshida}. After trapping of the atoms in the single beam optical dipole trap, evaporative cooling was performed at 300~G.
The magnetic field was ramped to the vicinity of the $\left\vert 1 \right\rangle-\left\vert 1 \right\rangle$ $p$-wave Feshbach resonance located at $B_0=159.17(5)$~G~\cite{inada}, which is consistent with the previous measurements $B_{0}^{\rm th}=159.15$~G~\cite{zhang, shunk}.
The error in the magnetic field $B$ is mainly arising from the magnetic fluctuation due to the noise in the current to create the magnetic field. We stabilize the current running in the coils to suppress the fluctuation down to $5 \times 10^{-5}$ which corresponds to a magnetic field fluctuation of 8~mG~\cite{naka}.

The magnetic field was initially tuned far above the $p$-wave Feshbach resonance $B_{0}$. Then, we abruptly ramped down the magnetic field to jump across the resonance toward the Bose-Einstein condensation side to avoid adiabatic creation of Feshbach molecules~\cite{inada}. 
Then, we quickly ramped the magnetic field to the various magnetic-field detunings $B-B_{0}$ and held the atoms in the trap to measure the loss of the atomic number.
As an example, Fig.~\ref{fig1}(a) shows the number of atoms for a hold time of 20~ms at various magnetic-field detunings. Here, the temperature of the atoms is $T \sim 1.5~\mu$K, and each data point is the average of three repeated measurements. The sharp edge of the loss feature in the lower magnetic-field side corresponds to the resonance position $B_0$~\cite{naka}. 
We treat the doublet structure of the $p$-wave Feshbach resonance due to the spin-dipole interaction~\cite{ticknor} as being completely overlapped with each other because the splitting of the two resonances in $^6$Li is quite small compared with the resonance width. 
The solid curve in Fig.~\ref{fig1}(a) shows the result of our theoretical description obtained using Eq.~(\ref{k3}) in the solution of rate equation.~(\ref{rate}), as discussed in a later part of the text.

To measure the three-body loss coefficient $L_{3}$, we monitored the hold-time dependence of the number of atoms $N$ at a fixed temperature $T$ and detuning from the Feshbach resonance $B-B_{0}$. The number of atoms decay according to the rate equation 
\begin{equation}
\frac{\dot{N}}{N}= - L_{3} \left\langle n^2 \right\rangle,
\label{rate}
\end{equation}
where $\left\langle n^{2} \right\rangle=(1/N) \int n(\vec{x})^{3}d\vec{x}$ is the mean square density determined from the atomic density profile obtained via absorption imaging. To simplify the analysis, we always keep the atomic temperature greater or equal to the Fermi temperature such that the density profile of atoms is assumed to be a Gaussian. 
We limited the atomic loss up to $30\%-35\%$ of the initial number of atoms, and the temperature change during the hold time is less than 10$\%$ as indicated in the inset of Fig.~\ref{fig1}(b). Therefore, we extracted the three-body loss coefficient $L_3$ from the solution of Eq.~(\ref{rate}) assuming a constant temperature approximation shown by the solid curve in Fig.~\ref{fig1}(b).
In our experiment, the one-body loss rate can be safely neglected because the one-body decay time of the atoms due to background-gas collision is 80~s and has negligible influence on the determination of $L_{3}$.
\begin{figure}[tbp]
\includegraphics[width=3.25 in]{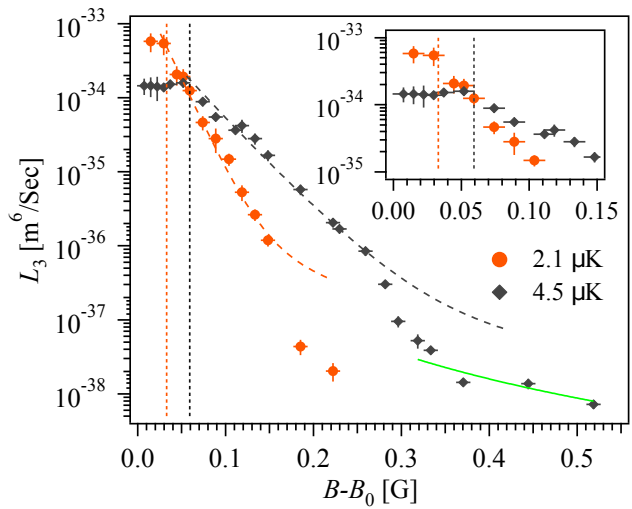}
\caption{Three-body loss rate coefficient $L_{3}$ as a function of magnetic-field detunings at two different temperatures of 2.1 (filled orange circles) and 4.5~$\mu$K (filled black diamonds). Vertical error bars show statistical errors due to the atom number fluctuations while horizontal error bars show 8~mG fluctuation of the magnetic field. When increasing the interaction, $L_{3}$ rapidly increases toward the resonance and hits the limit imposed by unitarity indicated by vertical dotted lines. The vertical dotted lines correspond to the universal value of $k_T / k_{\rm res} \approx 1$ (see text). $L_{3}$ then stays almost constant close to the Feshbach resonance, as clearly evident in the inset, which indicates unitarity-limited behavior. The dashed curves show the theoretical results of Eq.~(\ref{k3}) by taking into account the imaginary part of the scattering volume $V_{\rm i}$ as a single free parameter. In the weakly interacting regime, the solid green curve indicates the scaling law of $L_{3}\propto V^{8/3}$.}
\label{figb}
\end{figure}

Figure~\ref{figb} shows $L_3$ as a function of magnetic-field detuning on the negative-scattering-volume side of the resonance at temperatures of 2.1 (filled circles) and 4.5~$\mu$K (black diamonds). 
At a sufficiently far-off resonance regime, where interactions are weak, we marked the scaling of $L_{3}\propto V^{8/3}$ by the solid green line for the case of 4.5~$\mu$K, which was the subject of previous studies~\cite{suno, yoshida}.
The $L_{3}$ increases by orders of magnitude as the magnetic-field detuning decreases and eventually reaches the maximum limit imposed by the unitarity condition.
In the unitarity limit, $L_{3}$ shows no further dependence on the magnetic-field detunings toward the resonance which is clearly reflected by flat regions in the inset plot of Fig.~\ref{figb}.
These results of Fig.~\ref{figb} show a clear contrast to the case of an $s$-wave interacting unitary Bose gas, where the critical point of the magnetic field at which $L_3$ reaches the unitarity-limited value is independent of the atomic temperature~\cite{zoran}. 
We observed that $L_{3}$ starts to saturate toward the resonance at the universal value of $k_T / k_{\rm res}\geq 1$ as indicated by vertical dotted lines in the respective color in Fig.~\ref{figb}. Here, $k_T=\sqrt{3m k_{\rm B} T/(2\hbar^2)}$ and $k_{\rm res}=1/\sqrt{\left|V\right| k_{\rm e}}$,which are the thermal momentum and the momentum parametrized by the scattering volume $V$ and the effective range $k_{\rm e}$, respectively~\cite{russian}.

\begin{figure}[tbp]
\includegraphics[width=3.2 in]{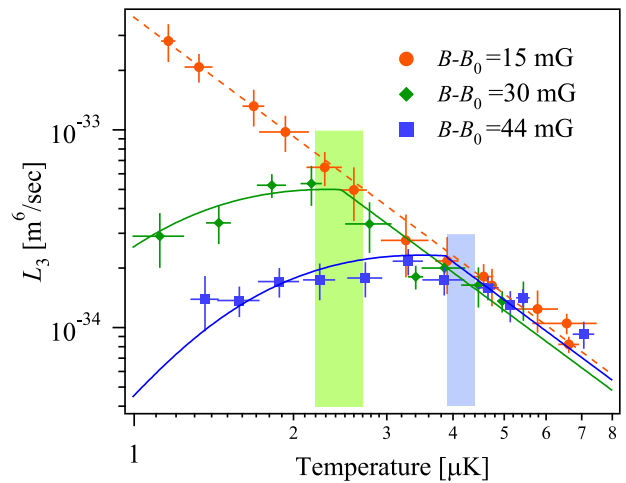}
\caption{The temperature dependence of $L_{3}$ for three different detunings: $B-B_{0}=15$ (orange circles), 30 (green diamonds) and 44~mG (blue squares). Each data point is the average of five repeated measurements under the same conditions. The error bars indicate the statistical error of 1~$\sigma$ standard deviation. Data at $B-B_{0}=15$~mG satisfied the universal unitary condition $k_T / k_{\rm res}\geq 1$ over the entire parameter range of the figure and followed the universal scaling law described by $L_{3} \propto T^{-2}$, as indicated by the orange dashed line. Data at 30 and 44~mG begin to deviate from the unitarity regime at the critical temperature region corresponding to the universal value of $k_T / k_{\rm res} \approx 1$ depicted by the vertical shaded band for each data set. The solid curves in the respective colors show the fitting result reproduced by Eq.~(\ref{unit}) in unitarity regime and by Eq.~(\ref{k3}) in the nonunitarity regime.}
\label{figunit}
\end{figure}

Another feature evident from Fig.~\ref{figb} is that the $L_{3}$ plateau at the unitarity regime is shifted to the lower value at the higher temperature. Therefore, we measured the temperature dependence of $L_{3}$ near zero detuning regions. Figure~\ref{figunit} shows the temperature dependence of $L_3$ at the unitarity-limited regime for three magnetic-field detunings; $B-B_{0}=15$ (orange circles), 30 (green diamonds), and 44~mG (blue squares). In this measurement, the atomic temperature was changed by controlling the degree of the evaporative cooling and the recompression condition of the dipole trap. Each data point is the average of five repeated measurements under the same experimental conditions. The error bars indicate 1~$\sigma$ standard deviation. 
In Fig.~\ref{figunit}, all the data points for $B-B_{0}=15$~mG satisfy the condition of the unitarity limit, $k_T / k_{\rm res}\geq 1$. At the unitarity limit, $L_{3}$ is expected to obey the following scaling law for the temperature~\cite{suno,suno2} 
\begin{equation}
L_{3}^{\rm max}=\lambda ~ \frac{36 \sqrt{3} \pi^{2} \hbar^{5}}{m^{3} (k_{\rm B} T)^{2}}.
\label{unit}
\end{equation}
Here, $\lambda \leq 1$ is a nonuniversal dimensionless constant whose unity value indicates the maximum upper bound of the three-body loss constant. The value of $\lambda$ is generally less than unity and is species dependent~\cite{zoran, green2}.
We fitted the data of $B-B_{0}=15$~mG with Eq.~(\ref{unit}) keeping $\lambda$ as free parameters.
The red dashed line shows the best fit result with $\lambda=0.09 \pm 0.02$, which confirmed the prediction of $L_3 \propto T^{-2}$.

The observed value of $\lambda$ indicates that the flow of the incoming hyper-radial wave into the scattering states of an atom and a deep dimer occurs with a probability less than 1.
The extracted value of $\lambda$ is different from those for Bose gases, which are 0.9 for $^7$Li ~\cite{rem} and 0.3 for $^{39}$K ~\cite{zoran}. As far as we know, there is no prediction of the value $\lambda$ for $p$-wave interacting $^6$Li atoms.

In the same way, the data taken at $B-B_{0}=30$ and 44~mG detuning show the scaling behavior at the relatively high-temperature regime, and the data start to deviate from the scaling behavior at low temperature. The vertical shaded bands in Fig.~\ref{figunit} indicate the temperature region where $k_T / k_{\rm res} \approx 1$. It is evident that the deviation of the data from the unitarity scaling behavior occurs at $k_T / k_{\rm res} \approx 1$ indicates the system is departing from the unitarity regime.

Next, we discuss the simplified theoretical interpretation for the line shapes of the atomic loss near the unitarity regime. Our scheme is based on the Breit-Wigner resonance scattering theory utilized to describe the laser-induced photoassociation trap loss~\cite{nap} and the $s$-wave three-body losses~\cite{juli,israel}. We assume that the two atoms with relative kinetic energy $E$ collide in the presence of a molecular bound state that is resonant with the collision energy. The two colliding atoms are then resonantly coupled to the quasibound state [i.e., $\rm Li+Li\rightleftharpoons Li_{2}(e)$]. The strength of the resonant coupling at collision energy $E$ exhibits a Wigner threshold form~\cite{victor}
\begin{equation}
\Gamma_{e}=\frac{2\sqrt{m} E^{3/2}}{k_{\rm e} \hbar}.
\label{ge}
\end{equation}
This threshold law also explains the lifetime of the quasibound molecular state~\cite{gab}, the dissociation energy of Feshbach molecules~\cite{waseem}, and the dynamics of contact relations~\cite{thy} on the negative side of $p$-wave Feshbach resonances. 
The atoms in the quasibound state undergo a relaxation into a deeply bound dimer $\rm Li_{2}(d)$ as a result of the collision with the third atom, which is represented by the relaxation event 
\begin{equation}
\rm Li_{2}(e) + Li \longrightarrow Li_{2}(d) +Li +\delta\epsilon,
\label{decay}
\end{equation}
with excess energy $\delta\epsilon$ added to the kinetic energy of the atom and dimer.

To obtain the cross section for the resonance-enhanced three-body recombination, we consider the effective range expansion of the $p$-wave scattering phase shift $\delta_{k}$ as $k^{3} \cot \delta_{k}= -1/V-k_{\rm e} k^{2}$, where $k$ is a relative wave vector and the effective range  $k_{\rm e}$ defined in this way is positive. In order to include the linewidth of the energy level associated with the dimer relaxation channels ($\rm d$) of Eq.~(\ref{decay}), we added the imaginary part $V_{\rm i}>0$ to the inverse of the scattering volume in the effective range expansion~\cite{russian,cpl,waseem2} 
\begin{equation}
k^{3} \cot \delta_{k}= -\frac{1}{V}-\frac{i}{V_{\rm i}}-k_{\rm e} k^{2}.
\label{range}
\end{equation}
Using standard $S$-matrix element notation~\cite{book1}, we derived the following expression for the cross section of resonance-enhanced three-body recombination due to atom-dimer relaxation~\cite{nap}
\begin{equation}
\sigma_{\rm p}=\frac{3 \pi}{k^{2}}~\frac{\Gamma_{\rm e} \Gamma_{\rm d}}{(E-E_{b})^{2}+(\Gamma_{\rm e}+\Gamma_{\rm d})^{2}/4}.
\end{equation}   
In this expression, $E_{b}=\hbar^{2} k_{\rm res}^{2}/m$ is the binding energy of the resonant quasibound state. The expression of the energy width $\Gamma_{\rm d}=2 \hbar^{2}/(m k_{\rm e} V_{\rm i})$ can be derived from the simple mathematics and $\Gamma_{\rm d}$ describes the rate of Eq.~(\ref{decay}). Therefore, $V_{\rm i}$ quantifies the strength of relaxation of the dimer into the deeply bound dimer states by collision with another atom.
The magnetic-field dependence of $V_{\rm i}$ has been discussed~\cite{russian} and it is predicted to be weak dependence on the magnetic field. In the magnetic-field range that we consider in this work, we can reproduce the experimental results by setting $V_{\rm i}$ to be independent of the magnetic field.
The three-body rate constant can be described as $K_{3}=(\phi_{\rm ae}/\phi_{\rm aaa}) v_{\rm rel} \sigma_{\rm p}$~\cite{marish}. Here, $v_{\rm rel}=2 \hbar k/m$ is the relative velocity and $\phi_{\rm ae}$ represents the phase space of an atom and resonant bound state while $\phi_{\rm aaa}$ is the phase space of three atoms. From $N$-particle phase space integral, one can obtain $\phi_{\rm ae}/\phi_{\rm aaa} \approx 96 \sqrt{3} \pi/k^{3}$~\cite{marish, brat}.
As a result, we obtain the three-body rate constant in the form of the Breit-Wigner expression given as
\begin{equation}
K_{3}=\frac{144 \sqrt{3} \pi^{2} \hbar^{5}}{m^{3} E^{2}}~\frac{4 \Gamma_{\rm e}\Gamma_{\rm d}}{(E-E_{b})^{2}+(\Gamma_{\rm e}+\Gamma_{\rm d})^{2}/4}.
\label{k3non}
\end{equation}
The measured value of the three-body loss coefficient $L_3$ is the thermal average of $l_{3}=3K_{3}/6$~\cite{suno} for an atomic ensemble. Therefore, we consider the thermal averaging over the Maxwell-Boltzmann distribution~\cite{ohara,qb}.
\begin{equation}
L_{3}= \frac{2}{\sqrt{\pi} (k_{\rm B} T)^{3/2}} \int_{0}^{\infty} l_{3} \times \sqrt{E}  e^{-E/k_{\rm B} T}dE .
\label{k3}
\end{equation}

We fit both the 2.1 and 4.5~$\mu$K data sets shown in Fig.~\ref{figb} with Eq.~(\ref{k3}) using $V_{\rm bg}\Delta B=(-2.8 \pm 0.3) \times 10^{6} a_{0}^{3}$~\cite{inada}, $k_{\rm e}\approx (0.091 \pm 0.01)~a_{0}^{-1}$, and $V_{\rm i}$ as the only free parameter.
Here, $a_{0}$ is the Bohr radius.
The van der Waals length scale of the $^6$Li atoms is $l_{\rm vdW} \sim 30 a_{0}$~\cite{dalgarno}, which can also be used to set the scaling in terms $l_{\rm vdW}$. 
Here, we can safely consider $k_{\rm e}=-\hbar^{2}/(m V_{\rm bg}\Delta B \delta\mu)$ because of the large scattering volume limit and narrowness of the resonance \cite{petrov, wang}, where $\delta\mu=k_{\rm B} (113 \pm 7)~\mu$K/G is the relative magnetic moment between the molecular state and the atomic state~\cite{fuch}. 
The dashed line curves in Fig.~\ref{figb} show the best-fit result with $V_{\rm i}=(6.5 \pm 1.5) \times 10^{-22}$~m$^3$. These theoretical curves explain the data in the intermediate region between the unitarity-limited regime and the weakly interacting regime.

Next, we fit the data shown in Fig.~\ref{figunit} with the combined curves of Eq.~(\ref{k3}) in the nonunitary regime and Eq.~(\ref{unit}) in the unitary regime. We take the critical temperature that divides the unitary and nonunitary regimes and the amplitude factors in both regimes as fitting parameters. The data agree well with the theoretical curves shown by the solid green and blue curves for $B-B_{0}=30$ and $B-B_{0}=44$~mG, respectively. The fitting errors of the critical temperature are shown with the width of the shaded bands in Fig.~\ref{figunit}. The temperatures corresponding to the condition of $k_T /k_{\rm res} \approx 1$ are consistent within the range of the fitting errors.
Similarly, Eq.~(\ref{k3}) is used in the solution of the rate equation (\ref{rate}) with the fixed trapping conditions~\cite{israel}, and a  theoretical curve is drawn for the number of atoms as the function of magnetic-field detunings. The solid blue curve shows the resultant theoretical curve in Fig.~\ref{fig1}, which is reasonably consistent with the experimental data.

The obtained value of the imaginary scattering volume $V_{\rm i}$ corresponds to the relaxation width of $\Gamma_{\rm d} \sim k_{\rm B} \times 0.15~\mu$K which is higher than the width of the resonant bound state $\Gamma_{\rm e} \sim k_{\rm B} \times 0.01~\mu$K and $\Gamma_{\rm e} \sim k_{\rm B} \times 0.04~\mu$K calculated from Eq.~(\ref{ge}) at $T= 2.1~\mu$K and $T= 4.5~\mu$K, respectively. 
This means that the timescale of the relaxation described by Eq.~(\ref{decay}) is shorter than the timescale of the association of the shallow Feshbach dimers scaled by $\Gamma_{e}^{-1}$. Therefore, the loss rate is dominantly set by the slow coupling of two atoms into the resonant quasibound state via barrier tunneling which is a two-body process at the current experimental conditions. Furthermore, $k_{\rm B} T \gg (\Gamma_{\rm e}+\Gamma_{\rm d})$ indicates the thermal broadening of the line shape of atomic losses as observed in the data of Fig.~\ref{figb}.

In summary, we experimentally investigated the unitarity-limited behavior of the three-body atomic loss near the $p$-wave Feshbach resonance in a single-component Fermi gas of $^6$Li atoms. We confirmed the universal scaling law of $L_{3}\propto T^{-2}$ at the unitarity limit, and we also confirmed that $k_T / k_{\rm res} \ge 1$ described the condition for the unitarity limit. The magnetic-field and temperature dependence of $L_{3}$ at the large negative scattering volume nearby the unitarity-limited regime shows good agreement with our theoretical description using the Breit-Wigner theoretical approach, which is based on the association of the resonant-bound state and its relaxation into a deep stable state by collision with another atom. Our approach directly relates the imaginary part of the inverse scattering volume to the relaxation lifetime of the Feshbach molecular bound state into deeply bound dimer states. 
Our result provides a promising direction to explore the few- and many-body physics at narrow unitarity regime using radio-frequency spectroscopy to minimize the effect of severe losses~\cite{thy}.

\emph{ACKNOWLEDGMENT}: The authors are thankful to G. V. Shlyapnikov and D. V. Kurlov for discussions.
This work was supported by a Grant-in-Aid for Scientific Research on Innovative Areas (Grant No. 24105006) and a Grant-in-Aid for Challenging Exploratory Research (Grant No. 17K18752). M.W. acknowledges the support of a Japanese government scholarship (MEXT).

\end{document}